# X20CoCrWMo10-9//Co$_3$O$_4$: a Metal-Ceramic Composite with Unique Efficiency Values for Water-Splitting in Neutral Regime

Helmut Schäfer*[a], Daniel M. Chevrier [b], Karsten Kuepper [c], Peng Zhang [b], Joachim Wollschlaeger,[c], Diemo Daum[d], Martin Steinhart [a], Claudia Heß [a], Ulrich Krupp[e], and Mercedes Schmidt [a]

[a] *Institute of Chemistry of New Materials and Center of Physics and Chemistry of New Materials, Universität Osnabrück, Barbarastrasse 7, 49076 Osnabrück, Germany*

[b] *Department of Chemistry, Dalhousie University, Halifax, Nova Scotia, Canada B3H 4J3*

[c] *Department of Physics, Universität Osnabrück, Barbaraßtraße 7, 49069 Osnabrück, Germany*

[d] *Faculty of Agricultural Science and Landscape Architecture, Laboratory of Plant Nutrition and Chemistry, Osnabrück University of Applied Sciences, Am Krümpel 31, 49090 Osnabrück, Germany*

[e] *Institute of Materials Design and Structural Integrity University of Applied Sciences Osnabrück, Albrechtstraße 30, 49076 Osnabrück, Germany*

**Abstract:** Water splitting allows the storage of solar energy into chemical bonds (H$_2$+O$_2$) and will help to implement the urgently needed replacement of limited available fossil fuels. Particularly in neutral environment electrochemically initiated water splitting suffers from low efficiency due to high overpotentials (η) caused by the anode. Electro-activation of X20CoCrWMo10-9, a Co-based tool steel resulted in a new composite material (X20CoCrWMo10-9//Co$_3$O$_4$) that catalyzes the anode half-cell reaction of water electrolysis with a so far-, unequalled effectiveness. The current density achieved with this new anode in pH 7 corrected 0.1 M phosphate buffer is over a wide range of η around 10 times higher compared to recently developed, up-to-date electrocatalysts and represents the benchmark performance advanced catalysts show in regimes that support water splitting significantly better than pH 7 medium. X20CoCrWMo10-9//Co$_3$O$_4$ exhibited electrocatalytic properties not only at pH 7, but also at pH 13, which is much superior to the ones of IrO$_2$-RuO$_2$, single-phase Co$_3$O$_4$- or Fe/Ni- based catalysts. Both XPS and FT-IR experiments unmasked Co$_3$O$_4$ as the dominating compound on the surface of the X20CoCrWMo10-9//Co$_3$O$_4$ composite. Upon a comprehensive dual beam FIB–SEM (focused ion beam-scanning electron microscopy) study we could show that the new composite does not exhibit a classical substrate-layer structure due to the intrinsic formation of the Co-enriched outer zone. This structural particularity is basically responsible for the outstanding electrocatalytic OER performance.

In spite of the drastic oil price collapse in the second half of 2014 resulting in a price below 30 $/barrel in January 2016 [1], the exploration of promising renewable energy sources for the

future is one of the significant challenges for scientists and engineers concerned with energy issues research. Splitting of water into hydrogen and oxygen by exploiting solar energy transforms water to an inexhaustible and environmental friendly fuel source [2, 3, 4, 5, 6, 7, 8, 9]. Electrocatalytically initiated hydrogen- and oxygen formation from water is considered an important realization of this solar to fuel conversion route [10, 11, 12] but is typically hampered by the high overpotentials oxygen evolution on the anode side goes with [13, 14]. This is particularly true when the electrochemical cleavage of more or less untreated water is intended-; hence, when the splitting procedure is carried out at neutral pH value. Noble metals like Pt, Ir, Ru, Au or noble metal containing compounds like $IrO_2$, $IrO_2$-$RuO_2$, $IrNiO_x$, $RuO_2$-NiO are famous for their comparatively low overpotential when used as an oxygen evolution reaction (OER) electrocatalyst in neutral regime [15, 16] but incur high acquisition costs. Therefore non-noble metal based OER electrocatalysts have been of interest for decades and the elaboration of the list of potential candidates suitable to ensure a stable and reasonable high oxygen evolution based current in neutral regime at potentials as close as possible to Nernstian potentials (1.228 V vs. the reversible hydrogen electrode (RHE) at standard temperature and pressure) is a striking field of activity [17, 18, 19, 20, 21, 22, 23, 24]. Scheme 1 gives some idea of the position of current heterogeneous catalysts in terms of their efficiency regarding OER in neutral regime.

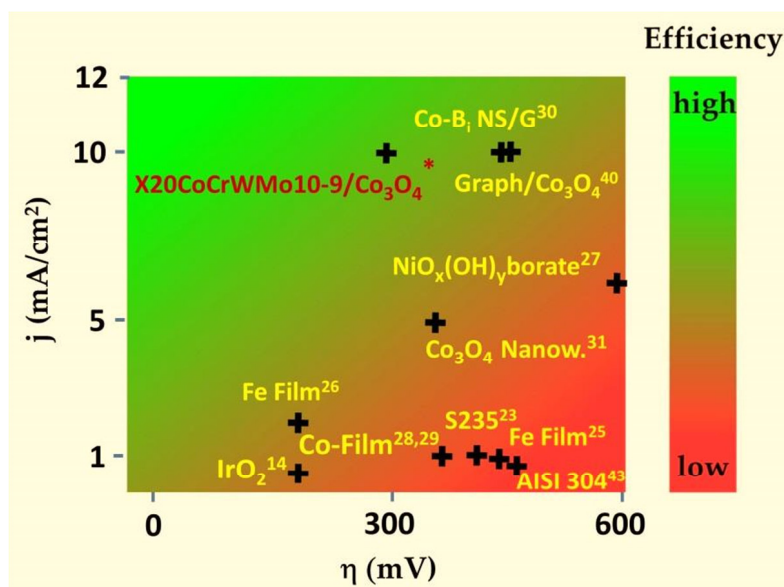

Scheme 1: The performance of recent developed heterogeneous catalysts for OER in neutral environment. * This work.

Very recently we reported on surface oxidized mild steel S 235 as prospective electrode material for the anodic splitting of water under neutral conditions. We determined for the OER upon iron oxide-based layers in 0.1 M phosphate (pH 7) buffer solution $\eta$=462 mV at 1 mA/cm² current density[25] which is still far below the OER performance of up to date electrocatalysts shown in alkaline regime. [26] Iron based films [27, 28], Ni-(oxy)hydroxide-borate [29], cobalt phosphate compounds [30, 31], cobalt borate/graphene [32], nano-scaled cobalt oxide based catalysts like $Co_3O_4$ nanowire arrays [33] and graphene $Co_3O_4$ nanocomposites [42] are among the non-noble-compounds for which a slightly better OER activity with overpotentials down to 492 mV at 10 mA cm$^{-2}$ has been demonstrated (Scheme 1). However, very rarely long-term voltage–current behavior or Faradaic efficiency was shown. In addition some of the protocols reported are laborious and precursors are costly. Thus motivated by the superlative-, sheer unrivalled properties of spinel $Co_3O_4$, which are limited in any case to OER electrocatalysis [34, 35, 36, 37, 38, 39] we intended to gain intrinsically grown-, $Co_3O_4$- based ceramic-alloy composite which has substantially better properties than the "dream-like" $Co_3O_4$ starting from simple steel. An outstanding electro-catalytically active and stable X20CoCrWMo10-9//$Co_3O_4$ composite was created via a straightforward electro-oxidation of-, a Co-based tool steel under harsh conditions.

**Results**

*OER properties in neutral medium*

Figure 1 represents a comparison of the electrochemical OER properties of X20CoCrWMo10-9 steel which had been anodized prior to electrocatalysis in 7.2 M NaOH at 1875 mA/cm² for 300 min, henceforth referred as sample Co-300, with untreated X20CoCrWMo10-9, designated as sample Co in pH 7 phosphate buffer solution. It should be mentioned at this point that <u>all</u> electrochemical measurements were carried out without any correction of the Ohmic voltage drop.

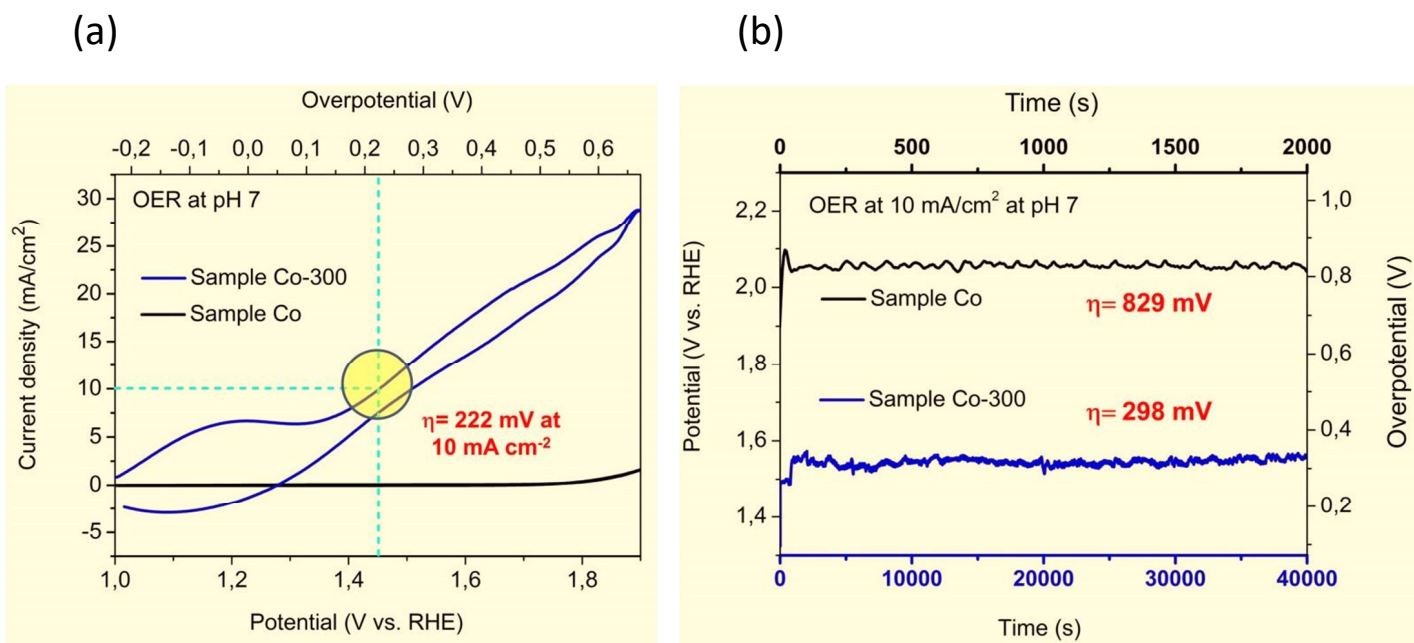

Figure 1. Overview of the electrochemical OER properties of non-treated X20CoCrWMo10-9 alloy (sample Co) and surface oxidized X20CoCrWMo10-9 (sample Co-300) in pH 7 corrected 0.1 M $KH_2PO_4/K_2HPO_4$. Electrode area of all samples: 2 $cm^2$. (a) Cyclic voltammetric plots of sample Co/sample Co-300 based on 20 mV/s scan rate and 2 mV step size. (b) Long term chronopotentiometry measurement of sample Co-300 and chronopotentiometry measurement of sample Co performed at 10 $mA/cm^2$ current density. Average overpotential for the OER through 40000 s plot: 298 mV (Co-300). Average overpotential for the OER through 2000 s plot: 829 mV (Co).

The significant improvement of the voltage-current behavior can be taken from both-, the non- steady state (Figure 1a) as well as the steady state polarization (Figure 1b) experiments. The CV of sample Co-300 shows along the entire curve substantially stronger current to voltage ratio than the CV of sample Co and reached, at the upper voltage limit of 1.9 V vs. RHE, an impressive current density of 28 $mA/cm^2$ (Figure 1a). A reasonable current density of 10 $mA/cm^2$ was achieved at only 1.45 V vs. RHE which corresponds to 222 mV overpotential (Figure 1a). Regarding the steady state performance in the neural regime, Co-300 fulfilled our expectations based on the outcome of the CV experiments. Oxidative water splitting was catalyzed outstandingly efficient and stable resulting in an unusually low overpotential (298 mV) required to realize 10 $mA/cm^2$ current density for 40000 s at pH 7 (Figure 1b). To the best of our knowledge such a strong current voltage ratio has never been shown before, neither in cyclic voltammetric-, nor in chronopotentiometry measurements performed at pH 7. Moreover, it represents the level of performance of state of the art electrocatalysts in regimes that support water splitting much better than pH 7 buffer solution like acids and bases. Thus electro-activated steel AISI 304 on which we reported recently exhibited a comparative behavior in 0.1 M KOH [26].

In general, most of the established OER electrocatalysts show at overpotentials of around 500 mV a OER current density that is approximately ten times lower (1 mA/cm$^2$) [18, 25, 27, 30, 40, 41, 45]. Thus, for instance, $Co_3O_4$ nanowire arrays developed by He *et al.* showed in polarization measurements [33] an unusually strong voltage-current behavior at pH 7.2 ($\eta$~480 mV at 5 mA cm$^{-2}$). The absence of steady state measurements performed at pH 7.2 and the absence of experiments quantifying the oxygen formation makes it difficult to conclusively assess the OER performance. Even slightly better properties have been shown for graphene-$Co_3O_4$ nanocomposites under neutral conditions (498 mV at 10 mA cm$^{-2}$) [42]. Unfortunately, no long term voltage-current behavior was shown and the release of oxygen was not quantified [42]. To confirm the assumption that the chosen electro-activation procedure creates, on the steel surface a passivating oxide layer that sufficiently protects the metal matrix below the layer against further oxidation and to confirm that the current determined upon electrochemical measurements is exclusively due to oxygen evolution, the oxygen formation was quantified and the mass loss determined. No mass loss occurring during long term operation at 10 mA/cm$^2$ and 5 mA/cm$^2$ was obtained (Table S1) and the charge to oxygen conversion rate after 2000 s operation time (Table S2, Figures 2a, 2b) amounted to 75.6% (10 mA/cm$^2$), and 83.2% (5 mA/cm$^2$); thus, reflecting a similar outcome when compared to findings we extracted from OER experiments with Ni, Fe-oxide coated AISI 304 steel in 0.1 M KOH [26]. The Faradaic efficiency of a variety of OER electrocatalysts under neutral conditions has been investigated by several groups but in most cases at significant lower current densities [31]. Manganese oxides based OER electrocatalysts as a model system for the oxygen evolving complex of photosystem II were frequently studied and based on head space measurements, the Faradaic efficiency amounted to 78% after 90 min of OER at lower current density in pH controlled $Na_2SO_4$ [43] solution. Low Faradaic efficiencies at higher current densities do not necessarily indicate an oxidation/dissolution of the catalysts itself but can up to some extent be caused by inaccurate measurements. We assume that at relative high current density, undissolved oxygen gas bubbles are generated, *i.e.* oxygen cannot be dissolved with the speed with which it is formed. Thus, a significantly lower

Faradaic efficiency by increasing current densities (From 97% at 1 mA/cm$^2$ to 43% at 10 mA/cm$^2$) was obtained by Qiu *et al.* (2014) for Ni-Fe containing nanoparticles in 1 M KOH [44].

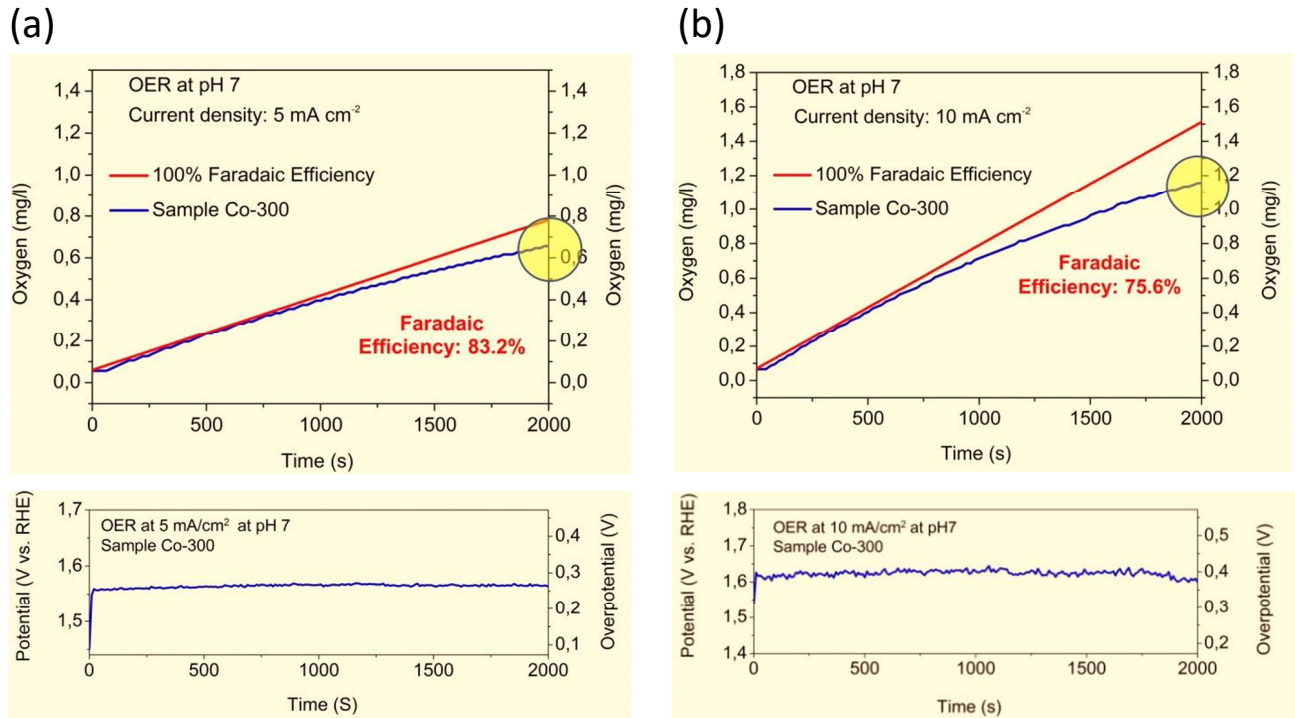

Figure 2. Faradaic Efficiency of the OER upon sample Co-300 at pH 7 whilst chronopotentiometry measurements (lower images) performed for 2000 s at 5 mA/cm$^2$ (a)-, 10 mA/cm$^2$ (b), respectively. Electrode area: 2 cm$^2$; Correlation of oxygen evolution upon sample Co-300 in 0.1 M K$_2$HPO$_4$/KH$_2$PO$_4$ (blue curve) with the charge passed through the electrode system (red line corresponds to 100% Faradaic efficiency). **(a)** Amount of the electrolyte: 2.3 l; Start value of dissolved oxygen: 0.06 mg/L (t= 0 s); End value of dissolved oxygen (t = 2000 s): 0.66 mg/L (nominal value (100%):0.781 mg/L). Faradaic efficiency of the OER after 2000 s runtime: 83.2 %. **(b)** Amount of the electrolyte: 2.3 l; Start value of dissolved oxygen: 0.07 mg/L (t= 0 s); End value of dissolved oxygen (t = 2000 s): 1.16 mg/L (nominal value (100%):1.51 mg/L). Faradaic efficiency of the OER after 2000 s runtime: 75.6 %.

A comparison of the OER properties of our steel samples at pH 7 with the corresponding characteristics of noble metal containing catalysts is a *sine qua non* for an in-depth evaluation of this material. We chose commercially available IrO$_2$-RuO$_2$ sputtered on titanium as comparison sample (sample Ir/Ru) for OER activity and stability at pH 7. A direct comparison of the OER performance of samples Co, Co-300 and Ir/Ru can be derived from Figure 3.

The CV curve of sample Co-300 (Figure 3a, blue curve) was found to be substantially stiffer than the CV curve of sample Ir/Ru, thus over the entire range of overpotentials sample Co-300 exhibited superior electrocatalytic OER activity when compared to Ir-Ru (Figure 3a, red curve). As expected, a similar outcome can be derived from the corresponding Tafel lines of samples Co-300 and Ir/Ru, showing a horizontal shift of around 220 mV (Figure 3b).This represents enormous progress, taking into consideration that our group had significant problems with substantially improving the OER properties of steel upon surface oxidation to a level that makes it competitive to $IrO_2$-$RuO_2$ regarding water-splitting properties under neutral conditions [25, 45].

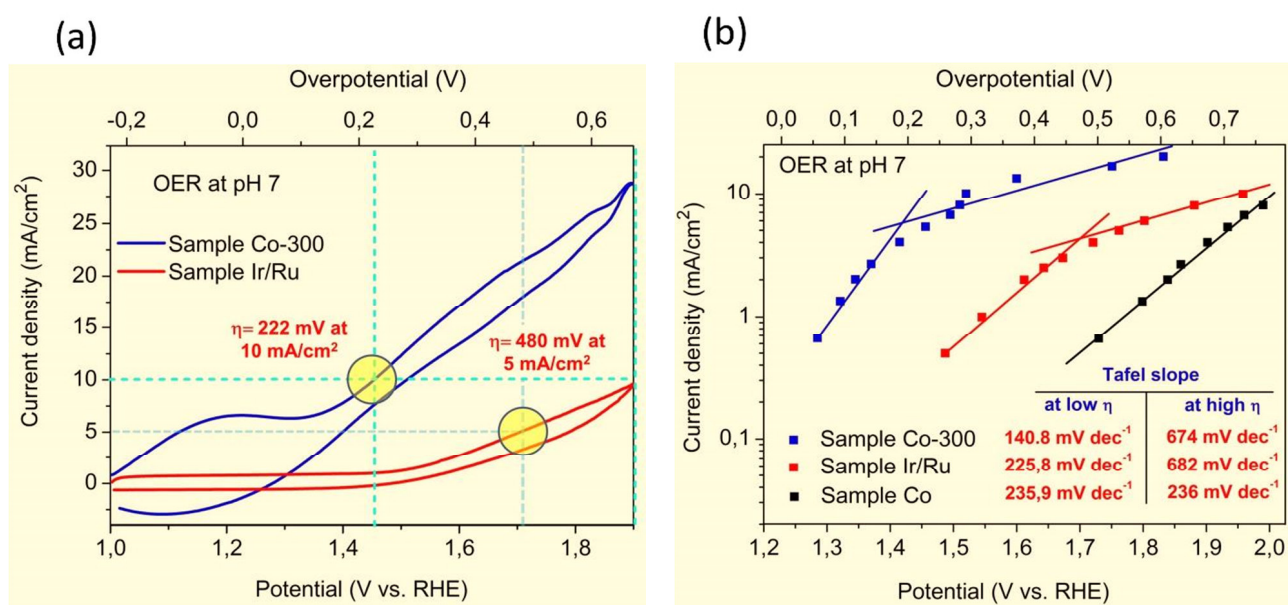

Figure 3. Comparison of the electrochemical OER properties of sample Co/sample Co-300 with sample Ir/Ru in pH 7 corrected 0.1 M $KH_2PO_4$/$K_2HPO_4$. Electrode area of all samples: 2 cm$^2$. (a) Cyclic voltammetric plots of sample Co-300 and sample Ir/Ru based on 20 mV/s scan rate and 2 mV step size. (b) Tafel plots of samples Co, Co-300 and Ir/Ru based on 200 second chronopotentiometry scans at current densities 0.65, 1.33, 2, 2.66, 4, 5.33, 6.66, 8, 10, 13.3, 16.6 and 20 mA/cm$^2$.

A substantial horizontal shift (~450 mV) of the Tafel line of sample Co-300 (blue squares) compared to the corresponding Tafel line of untreated steel Co (black squares) towards lower potentials (Figure 3b) proved what we already derived from Figure 1: the meaningful enhancement of the OER relevant electrocatalytic properties under neutral conditions upon the applied surface oxidation. This horizontal shift between Tafel lines assigned to the Co sample and the Co-300 sample can only be explained by a change in the chemical nature of the surface during surface oxidation whereas an increase of active surface during the surface oxidation would lead to a vertical shift of CV curve or Tafel line.

AFM experiments done with Co/Co-300 (Figures S1) helped to further exclude this kind of strong "surface area" effects.

Interestingly, at current densities < 5mA/cm$^2$ the Tafel line belonging to sample Co-300 was substantially stiffer (slope 140.8 mV dec$^{-1}$) than the one assigned to IrO$_2$-RuO$_2$ (slope 225.8 mV dec$^{-1}$) (Figure 3b). The Tafel plot of sample Co-300 showed some signs of a slight dual Tafel behavior with lower slope at lower overpotential region (140.8 mV dec$^{-1}$) and higher slope at higher overpotential region (slope 674 mV dec$^{-1}$). In principle the dual Tafel characteristics could be a kind of artifact due to the lack of voltage drop compensation (IR compensation), where R is the sum of contributions of both electrolyte and electrode resistances. We therefore performed an IR compensation on the polarization measurements the Tafel plots are based on in order to achieve a more reliable testimony (Figure S2 new). As also obtained by other groups[46] IR compensation via standardized software was found to be too strong, i.e. lead to anomalous voltage–current behavior (reduced current at increased potential) in the high potential region, we corrected Ohmic losses as shown in one of our previous reports[25] manually by subtracting the Ohmic voltage drop from the measured potential on the basis of an electrolyte (pH 7 buffer solution) resistance of 3.4 Ω at 0.5 mm electrode distance [47]. Dual Tafel behavior was found to be less pronounced but remained for samples Co-300 and Ir/Ru (Figure S2 new, blue and red curve). Independently from the correction of the voltage drop untreated steel X20CoCrWMo10-9 exhibited single Tafel behavior with a slope of 181 mV dec$^{-1}$ under neutral conditions for the IR corrected data (Figure S2 new, black curve), a slope of 235.9 mV dec$^{-1}$ for the non-IR corrected data respectively (Figure 3b, black curve). Due to the fact that this single Tafel behavior occurs at quite high overpotential we assume that the dual Tafel behavior obtained for samples Co-300 and Ir/Ru at significant lower potential does not simply indicating the bottleneck effect due to mass transport limitations. The onset of diffusion limitation of the electrocatalytic OER depends on many parameter but is considered to occur not below 20 mA/cm$^2$ [48].

Significantly reduced Tafel slopes were obtained for all samples when IR corrected data were used with for instance a slope of 97.8 mV dec$^{-1}$ determined for Co-300 in the lower overpotential- and 291.1 mV dec$^{-1}$ determined in the higher overpotential region (Figure S2 new, blue curve). This agrees very well with values determined recently for the OER upon

graphene-$Co_3O_4$ nanocomposites in 0.1 M phosphate buffer solution at lower overpotentials (98 mV dec$^{-1}$) [42]. Evolving oxygen bubbles may block electrochemical active sites as obtained by Lu et al. [49] and could in addition cause a higher Tafel slope at higher potential, i.e. a transition between regions with different slopes. Besides these "technical effects" dual Tafel characteristics could imply different OER mechanism, different catalytically active species, or a different rate-determining step for the OER at different overpotentials upon the surface of Co-300 [50]. Of course increased Tafel slopes at increased potentials might also be a sign of degradation of the catalyst but can be excluded here in case of all three samples tested here (Figures 3b, S2 new) as repeated determination of the Tafel plots did not lead to other findings.

*The origin of the high catalytic activity*

In our recent report we unmasked a dissolution process to be likely responsible for the enrichment of catalytic active species on the surface of AISI 304 steel during electro-activation[26]. A substantial mass loss of around 6 mg took place whilst the electro-activation of X20CoCrWMo10-9 steel as well and in addition the formation of a black-grey layer on the Pt-counter electrode could be obtained (Table S3). These findings clearly prove parallels regarding the process of a hypothetical layer formation is based on when AISI 304 and X20CoCrWMo10-9 steel are electro-oxidized in strong alkaline medium. We investigated the electrolyte (7.2 M NaOH), as well as precipitation formed on the counter electrode during altogether five electro-activation procedures of X20CoCrWMo10-9 steel via ICP OES. The results can be taken in detail from the supporting information (Table S4). During activation the steel significantly lost Fe and Cr as well as Mo. Interestingly no substantial amounts of Co could be detected either in the electrolyte used for the electro-activation or in the layer formed on the counter electrode (Table S4). Thus, based on the outcome of these investigations, we assumed the formation of a Co-oxide layer on the surface of X20CoCrWMo10-9 steel upon hard anodization. Obviously a drastic enrichment of Co via suppression of Fe, Cr, W and Mo on the periphery of the Co-based tool steel had taken place at strong oxidative potential in alkaline solution. From a XPS study performed with sample Co and Co-300 after carrying out long term polarization experiments we hoped to manifest this assumption and we wanted to identify which species are present on the surface of untreated and the oxidized X20CoCrWMo10-9 steel.

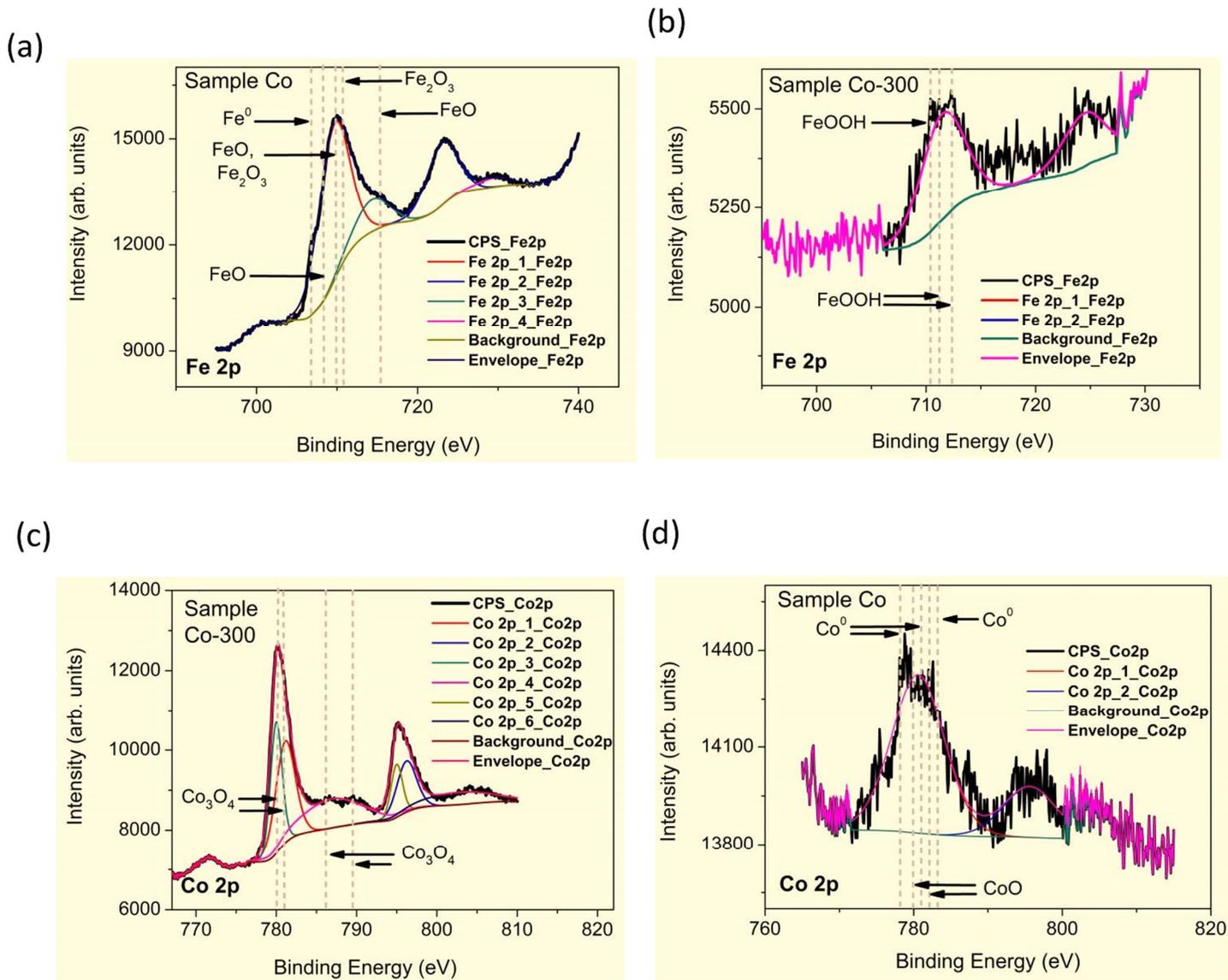

Figure 4. High resolution XPS spectra of samples Co and Co-300. Binding energies of reference compounds [51] are indicated by vertical lines as visual aid. **(a)** Fe 2p core level spectra. Fitting results for sample Co at peak positions 709.619 eV, 714.323 eV, 723.016 eV and 729.536 eV. **(b)** Fe 2p core level spectra. Fitting results for sample Co-300 at peak positions 711.452 eV and 724.416 eV.
**(c)** Co 2p core level spectra. Fitting results for sample Co-300 at peak positions 779.98 eV, 781.101 eV, 786.922 eV, 794.982 eV, 796.201 eV and 802.78 eV. **(d)** Co 2p core level spectra. Fitting results for sample Co at peak positions 780.603 eV and 795.624 eV.

A cursory glance at Figure 4 is sufficient to detect the huge differences of samples Co and Co-300 regarding the Co and Fe content on the surface. Low Co content on the surface of Co and low Fe content on the surface of Co-300 can already be revealed from low signal to noise ratio in the high resolution Co-2p and Fe-2p XPS spectra (Figures 4b, d). Indeed, while Fe is the predominant metal element on sample Co (63.5 at %), Co clearly dominated on the surface of sample Co-300 (71.6%, Table S5), which impressively exhibited the enrichment of Co/suppression of Fe on the surface of X20CoCrWMo10-9 steel during electro-oxidation. Besides Fe, Cr was also found to be depleted on the surface upon hard anodization (Figure S2, Table S5). The Co $2p_{3/2}$ binding energies and the overall peak shape of the Co 2p core level spectra of sample Co-300 (Figure 4c) agree very well with the corresponding data derived from the XPS analysis of $Co_3O_4$ reported by Biesinger et al [51]. The Co $2p_{3/2}$ related peaks are located at binding energies of 780 eV, and 786.4 eV, and the Co $2p_{1/2}$ peak is located at 795.2 eV. The charge transfer satellite which is characteristic for $Co_3O_4$ is also visible around 789.5 eV. Based on these findings we intend to unmask spinel type $Co_3O_4$ as the catalytic active compound on the surface of Co-300. Regarding the overall catalytic activity we consider the impact of Mn oxide species, an admixture in the layer on Co-300 to be low (Table S5, Figure S2). The same should apply to iron oxide compounds (very likely FeOOH) which could be detected in traces on the surface of Co-300 (Figure 4b). Details of the outcome from the XPS study performed with samples and Co-300 can be taken from the supporting information (Figure S2).

Binding energies of the corresponding reference compounds were extracted from literature [51, 52, 53]. Herewith, the discussion should be limited to the finding that all relevant alloying ingredients, such as Fe, Co, Cr, Mn, Mo, and W could be detected on the surface of untreated steel X20CoCrWMo10-9 (Figures S2c, d, e, h, Table S5) whereas, for instance, W, Mo and Cr were completely suppressed on or close to the surface of sample Co-300 (Table S5). Due to the electrochemical experiments in KOH performed with our samples prior to these XPS study the surface was contaminated with K and P (Figure S2g).

A FTIR spectroscopic study carried out with sample Co-300 after completion of long term polarization tests should be sufficient to obtain direct evidence for the existence of molecular $Co_3O_4$ on the surface of electro-activated X20CoCrWMo10-9. Sample Co-300 has been used for this study IR-measurements were performed in ATR (Attenuated Total Reflection) mode as this technique allows the direct detection of molecular species in the surface of solids. Unfortunately we could not extract a meaningful FTIR spectrum from the oxide containing surface of Co-300 upon direct measurement of the specimen. This experience we had already put before when surface oxidized steel S235 or AISI 304 were studied [25, 26, 45]. As the evanescent wave penetrates only up to some microns into the sample a good contact between ATR diamond and the sample is crucial and obviously was not ensured. Therefore the active outer sphere of Co-300 was mechanically removed and the resulting powder was used for the FTIR study. $Co_3O_4$ is known to show two absorption bands with medium amplitude at specific positions (~550 and 650 $cm^{-1}$) due to stretching vibration of the metal-O bond [54, 55, 56]. Indeed, these absorption bands can be clearly seen in our spectrum (Figure S3) indicating that Co is octahedral surrounded by oxygen. This finding represents the desired proof for the structural confirmation of $Co_3O_4$. Absorption bands positioned at somewhat higher wavenumber suggest also the existence of oxide-, or hydroxide oxide species (Figure S3). The outcome from the XPS- and the IR studies gives us some idea why sample Co-300 shows this enormous activity towards catalytically initiated OER at pH 7 (Figures 1-3). However, it should not go unmentioned that other, for instance upon electrodepostion generated $Co_3O_4$ based OER electrocatalysts were found to be substantially less active [33, 42] than Co-300, which raises the suspicion that the existence of $Co_3O_4$ on the top of our electro-oxidized steel is a necessary but not a sufficient requirement for our findings. We assumed that the intrinsic formation of the $Co_3O_4$ based outer sphere of the electro-activated steel is besides the generally high catalytic activity of $Co_3O_4$ as such responsible for the very sufficient initiation of OER upon Co-300. Thus, we expected a lower catalytic activity in case that X20CoCrWMo10-9 steel was coated by a $Co_3O_4$ layer, *i.e.* when the composition $Co_3O_4$ on the periphery of the sample was not implemented from the "inside of the material itself" but through a conventional coating process. We decided to perform an additional experiment in order to provide more valuable insights regarding this issue. X20CoCrWMo10-9 steel was coated with a $Co_3O_4$ layer designated as sample Depos-30.

Sample Depos-30 was generated using a twostep electrochemical approach consisting of cathodic electrodeposition of $Co(OH)_2$ on X20CoCrWMo10-9 substrate (Figure 5a, pink curve) similar to the procedure described in literature [57] subsequently followed by electrochemical oxidation of the thin layer upon chronopotentiometry performed at 10

mA/cm$^2$ for 4000 s (Figure 5a, blue curve). XPS analysis of the surface of sample Depos-30 confirmed the expectation that the layer basically consists of Co$_3$O$_4$ (Figure 5c).

The Co 2p$_{3/2}$ binding energies and the overall peak shape of the Co 2p core level spectra of sample Depos-30 (Figure 5c) agrees perfectly with sample Co-300 (Figure 4c). The Co 2p$_{3/2}$ peak is located at a binding energy of 780.5 eV, at 785.4 eV respectively. The Co 2p$_{1/2}$ peak is located at 795.8 eV. The charge transfer satellite which is characteristic for Co$_3$O$_4$ is also visible at around 790 eV. Despite comparable surface composition, Depos-30 was found to be substantially less active toward electrochemically initiated oxygen formation in pH 7 regime than Co-300 as can be seen in the corresponding CV plots (Figure 5b, green curve), and-, the chronopotentiometry plot (Figure 5a, blue curve). The reader should keep in mind that η=298 mV was required to initiate a stable OER current density of 10 mA/cm at pH 7 upon the surface of Co-300 (Figure 1b). Over the entire potential range Co-300 exhibited considerably higher current density than Depos-30 which lines up perfectly in between sample Co-300 and sample Co (blue, green and black curve, Figure 5b). Notably: An overpotential of around 500 mV was required to ensure a constant OER current density of 10 mA/cm$^2$ for sample Depos-30 in pH 7 corrected buffer solution (blue curve, Figure 5a) that agrees very well with the overall OER activity found for graphene/Co$_3$O$_4$ by Zhao *et al.* [42] in the neutral pH regime. In addition, the non-steady state voltage current behavior of Depos-30 (h = 495 mV at 5 mA/cm$^2$, Figure 5b, green curve) at pH 7 is equal to what has been reported for Co$_3$O$_4$ nano-wire arrays at pH 7.2 by He *et al.* (η = 480 mV at 5 mA/cm$^2$ [33]). Thus, although the composition of the surface of Depos-30 and Co-300 does not differ significantly, the OER performance does and the high catalytic activity of sample Co-300 cannot be solely explained by the fact that it`s "outer sphere" consists basically of Co$_3$O$_4$. Since the base material of both samples is the same (X20CoCrWMo10-9) it suggests that the conditions in the gap between substrate and the surface play a major role in the ability of the material to act as a good OER electrocatalyst. In addition, the very often obtained "island like" growth of electrodeposited layers in combination with a covering degree significantly below 100% could also be responsible for the lower OER performance of sample Depos-30. Nocera and co-workers for instance investigated on ITO electrodeposited Co based electrocatalyst and could upon SEM still see the ITO substrate through cracks in the film [30]. In contrast to these findings the surface of sample Co-300 proved to be compact and homogeneous. As expected the surface of the alloy was completely oxidized and the bare metal cannot be seen after 300 min of electro-oxidation under harsh conditions (Figure 6a). There is no contrast in the SEM image of sample Co-300 achieved with the back scatter detector (BSD) (Figure 6b) clearly showing that there are no areas with different composition. The finding was completely different when a Co-based oxide layer on the surface of X20CoCrWMo10-9 steel was created upon electrodeposition (Figure 6 c-f). Remarkable differences regarding the composition of the surface of Depos-30 can be revealed from the contrast in SEM images whenever a BSD detector was used (Figure 6 d, e). The longitudinal grooves of the bare steel surface caused by the grinding pretreatment can be clearly seen in the SEM image 6f. Moreover, it turned out that the electrodeposited layer is incompact and exhibited cracks (Figure 6e). It is reasonable to assume that the significantly lower OER electrocatalytic activity of sample Depos-30 when compared to Co-300 is due to lower covering degree and the cracks of the deposited layer (Figure 6 c, f).

Cross sectional analysis (vertical plane imaging) of samples Depos-30 and Co-300 was performed by dual beam FIB–SEM technique in order to get detailed information about the specific gap between substrate and the surface and to exhibit the expected "intrinsic" differences between Depos-30 and Co-300. The focused ion beam was used to cut out a trapezoidal trough in the surface deep enough to cover the complete outer zone of the specimen in which the composition and the microstructure of the material varies. Figures 7a-d show SEM images of the rear wall of the trapezoidal trough thus representing cross sections of Co-300 (a, b), and Depos-30 (c, d). Notably: the low contrast in SEM image 7a of sample Co-300 recorded with an energy selective backscattered detector (ESB) suggests that there is no abrupt change in the composition of Co-300 along the area investigated and the metal-ceramic transition can hardly be seen. Energy dispersive X-ray spectroscopy (EDS) was carried out and the spectra taken along the FIB machined cross section of Co-300 (Figure S4a) confirmed this quantitatively (Figure S4b-f). The Cr- and Mn concentration was below the detection limit. Closer to the surface iron was suppressed whereas the Co, K, P, O content was found to be increased in the outer zone (Figure S4 b-f), which is in perfect agreement with the XPS results (Figure 4, S2, Table S5). The increment of K and P content towards the surface can be easily explained by the electrochemical measurements carried out with samples Co-300 and Depos-30 in KOH and $PO_4^{3-}$ electrolyte prior to the FIB-SEM and the EDS study. The concentration of Co and O in the surface of Co-300 increased almost linearly by ~ four times (seven times) along a distance of 8 µm towards the surface (Figure S4 b, d). The outer $Co_3O_4$ zone of Co-300 was found to be compact in a way that there is no access anymore from the outside to the substrate (Figure 7a, 7b), *i.e.* there are no vertical tunnels which agrees very well with the plan view SEM images taken from Co-300 (Figure 6 a, b). A substrate-layer structure cannot be assigned to electro-oxidized steel X20CoCrWMo10-9 as can be taken from the SEM image achieved with a secondary detector (SESI. Figure 7b) as well. It is reasonable to assume that this specific microstructure is caused during the electro-oxidation upon harsh conditions. In deep contrast to sample Co-300, the cross section of sample Depos-30 exhibited a classical substrate-layer structure (Figure 7c, d) with a sharp metal-ceramic transition (Figure 7c) and an abrupt change in the composition of the material at the substrate-layer junction (Figure 7c). Furthermore, the cracks seen in the plan view images (Figure 6e) could be confirmed (Figure 7d) and definitely allow the electrolyte to access the bare metal during OER experiments which will certainly reduce the OER performance. The rapid change of the composition could indeed be confirmed by the EDS spectra taken along the cross section of Depos-30 as well (Figure S5). Especially the Fe- and the Co content was found to be rapidly decreased (increased) at the substrate-layer junction towards the surface of the specimen (Figure S5 e, f). In the case of sample Co-300 the concentration of Co and O changed by ~ four times (seven times) along a distance of 8 µm towards the surface (Figure S4b, d), the classic substrate-layer structure of Depos-30 exhibited a change of the Co (O) concentration by three times (nine times) along a distance of only two µm towards the surface (Figure S5b, e).

Thus starting from the same steel-substrate the electro-oxidation-approach resulted upon an intrinsic growth process in a composite material (Co-300) with a complete different microstructure and substantially better electrocatalytic OER properties than the electrodeposition approach (Depos-30) did.

*OER properties in alkaline medium*
Multiple pH characteristics of electrode materials will definitively support the widespread application of these materials. In addition to experiments performed under neutral conditions, we checked the suitability of surface oxidized X20CoCrWMo10-9 as an OER electrocatalyst in 0.1 M KOH as well (Figure 8). Again sample Co-300 showed superior oxygen evolution properties when compared to sample Co under both, steady state- and non-steady state conditions. The significant horizontal shift of the CV of sample Co-300 towards lower potentials by more than 300 mV relatively to the CV of sample Co can be clearly seen in Figure 8a. Sample Co-300 exhibited an impressive current density of 50 mA/cm$^2$ at 1.7 V vs. RHE (Figure 8a) that exceeded the current density/voltage ratio of electro-activated AISI 304 [26] and 316L steel[58] and is superior to the OER performance of most catalysts recently developed [13, 59]. Outstanding current-voltage performance was proved in low potential regions as well: An ultralow overpotential < 100 mV resulted in j= 10 mA/cm$^2$ (Figure 8a). $Ni_3S_2$ nanorods developed by Zhou *et al.* showed at $\eta$ =157 mV onset of OER[60]. Data derived from chronopotentiometry measurements also confirmed that sample Co-300 is a fantastic OER electrocatalyst in 0.1 M KOH: The average overpotential amounted to 230 mV whilst 40 ks of OER at j= 10 mA/cm$^2$ (Figure 8b) whereas activated AISI 304 required under the same condition $\eta$=269.2 mV [26].

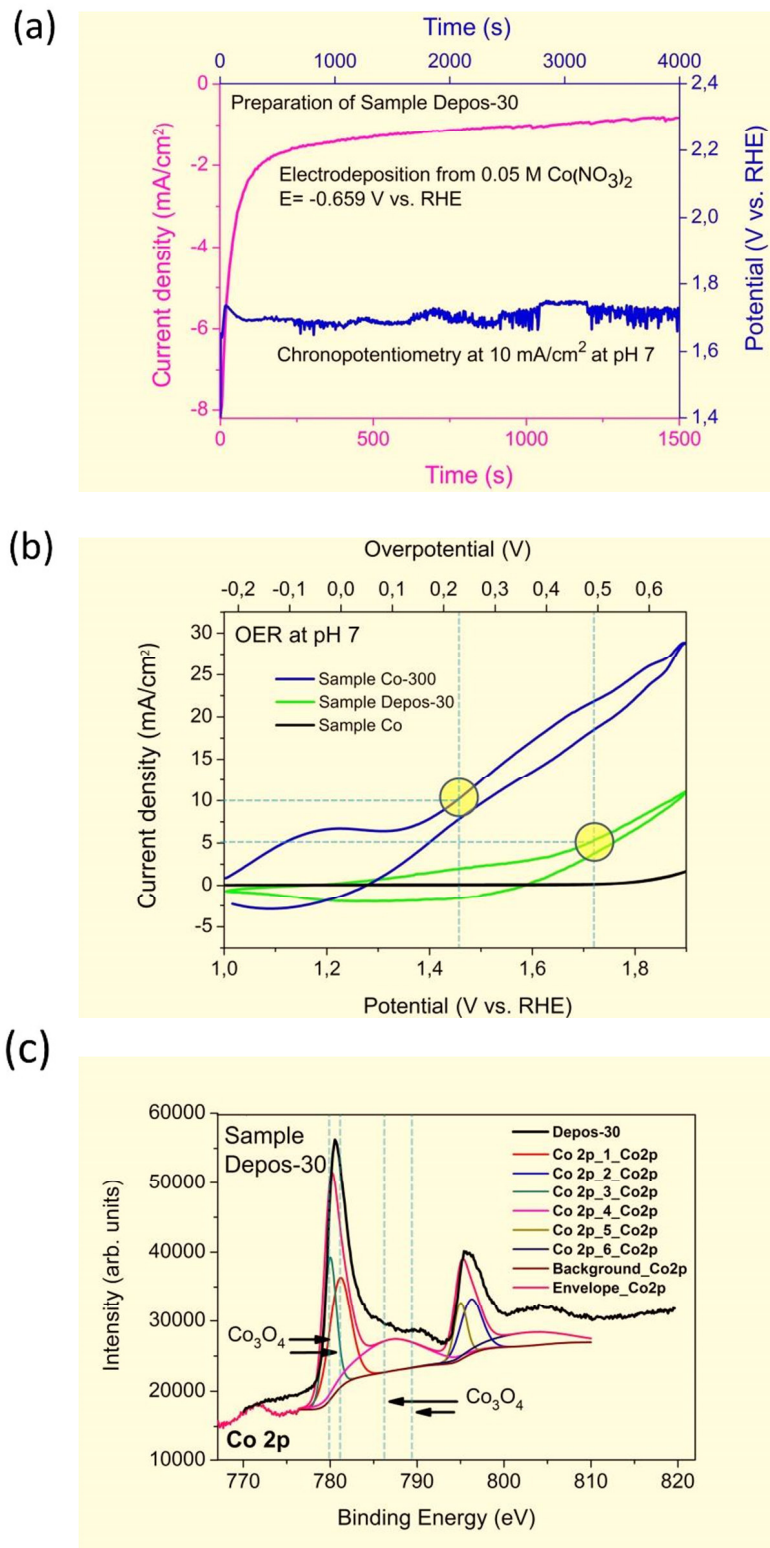

Figure 5a. Twostep (I+II) electrochemical preparation of sample Depos-30 upon electrodeposition from 0.05 M Co(II) nitrate solution. Electrode area: 3.255 cm2; Amount of electrolyte: 80 mL. (I) Current density/time dependence whilst cathodic deposition at constant potential of -0.659 V vs. RHE (magenta curve). (II) Chronopotentiometry plot of the as prepared sample carried at 10 mA/cm2 current density in pH 7 corrected 0.1 M $KH_2PO_4/K_2HPO_4$ buffer solution. **Figure 5b**. Comparison of the non-steady state electrochemical OER properties of samples Co, Co-300 and Depos-30 in pH 7 corrected 0.1 M $H_2PO_4/K_2HPO_4$. Electrode area of all samples: 2 cm$^2$. Cyclic voltammetric plots of samples Co, Co-300 and Depos-30 based on 20 mV/s scan rate and 2 mV step size. **Figure 5c**. High resolution XPS spectra of samples Co and Co-300. Binding energies of reference compounds [51] are indicated by vertical lines as visual aid Co 2p core level spectra. Fitting results for sample Depos-30 at peak positions 779.98 eV, 781.101 eV, 786.922 eV, 794.982 eV, 796.201 eV and 802.78 eV

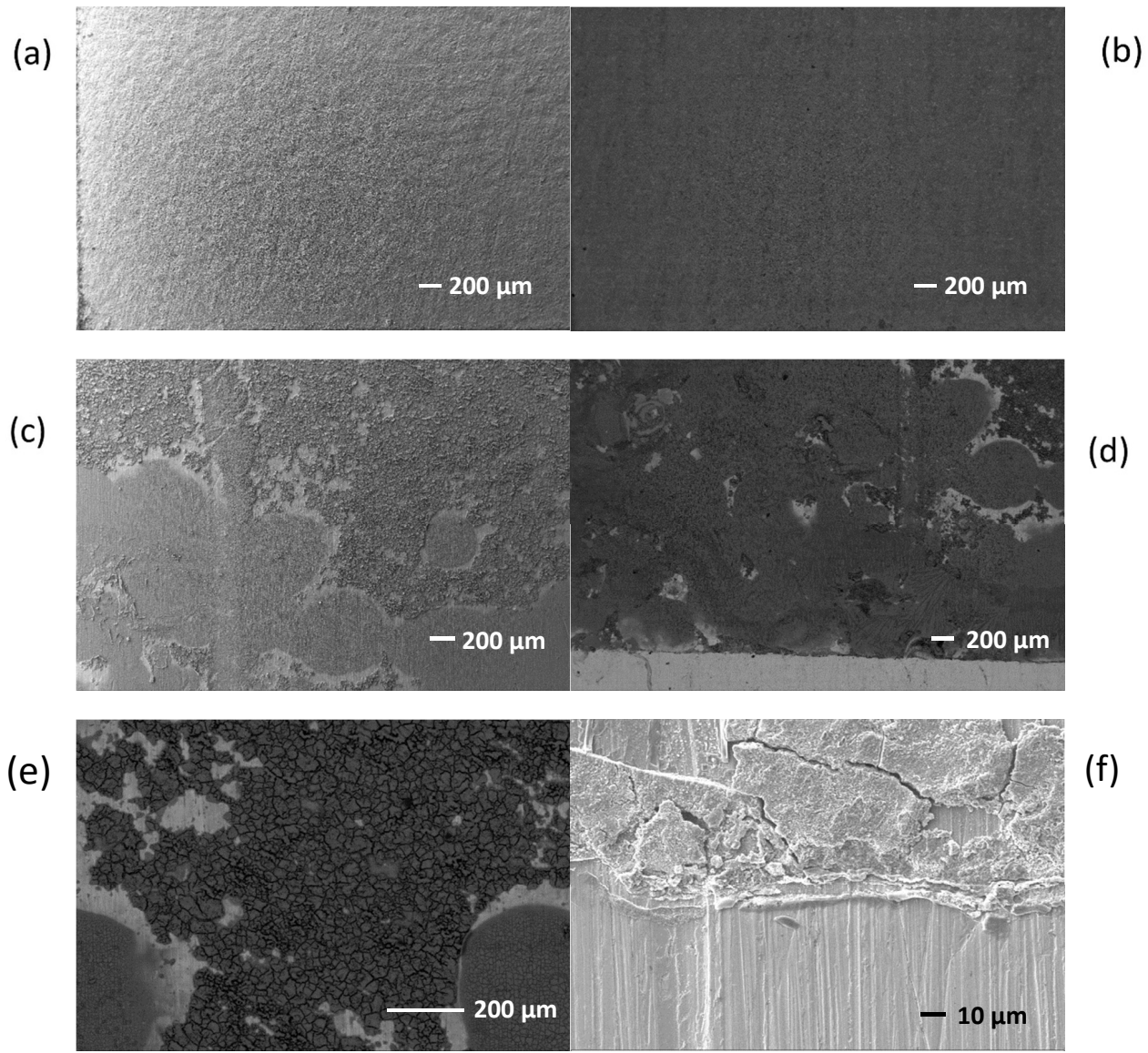

Figure 6. SEM images showing the topography of sample Co-300 (a, b), sample Depos-30 (c, d, e, f) respectively. The accelerating voltage was adjusted to 10 kV and the SEM images were acquired with a secondary electron detector SESI (a, c, f), back scatter detector NTS BSD (b, d, e) respectively. NTS BSD detector is used for qualitative compositional imaging. Areas of the sample with heavier elements appear brighter. Ion (Ga) beam settings: current: 240 pA, voltage: 10 kV, duration: 25 min.

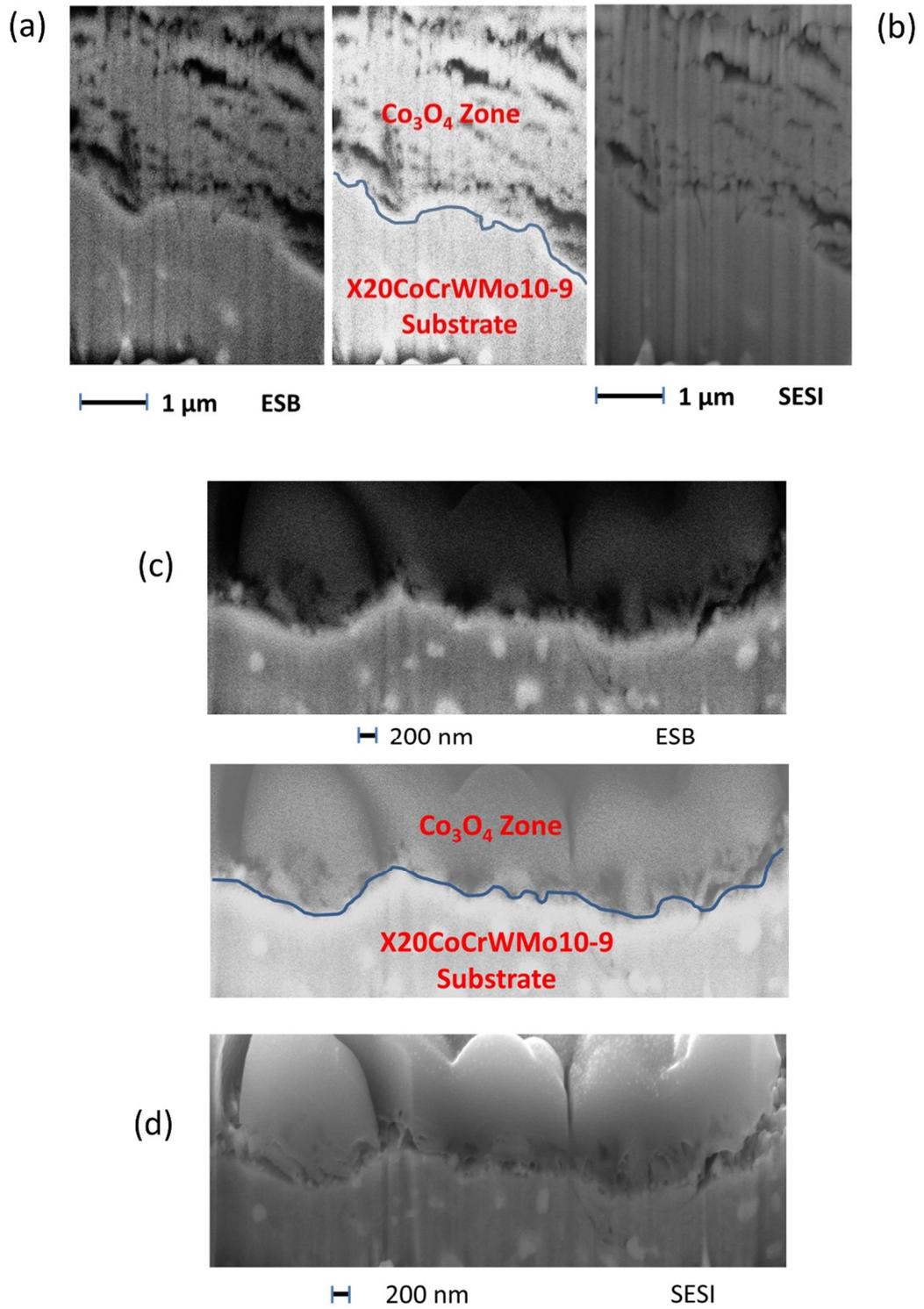

Figure 7. SEM micrograph of a FIB machined cross section of sample Co-300 (a, b), sample Depos-30 respectively (c, d). Shown is the rear wall of the trapezoidal trough. This wall is orientated perpendicular to the surface of the specimen thus presenting a cross section of samples Co-300 and Depos-30. The accelerating voltage was adjusted to 10 kV and the SEM images were acquired with a secondary electron detector SESI (b, d), Energy selective Backscattered detector ESB (a, c) respectively. Ion (Ga) beam settings: current: 240 pA, voltage: 10 kV, duration: 25 min.

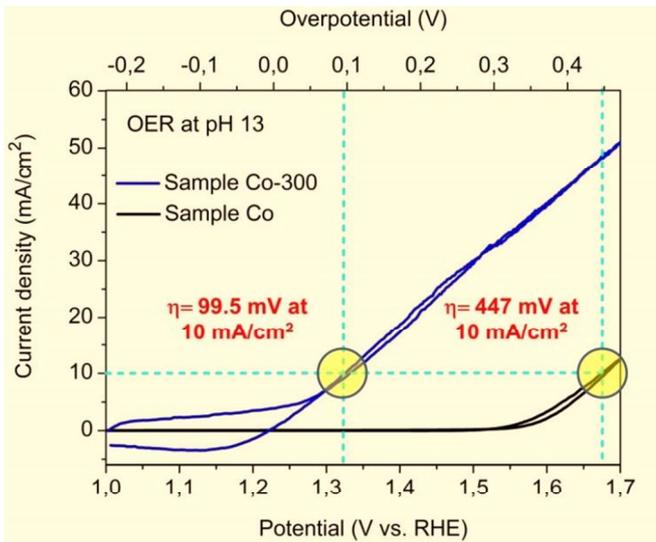
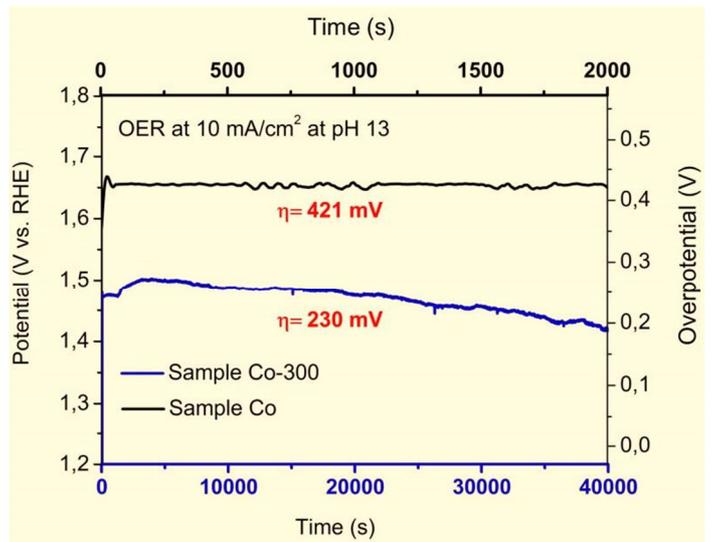
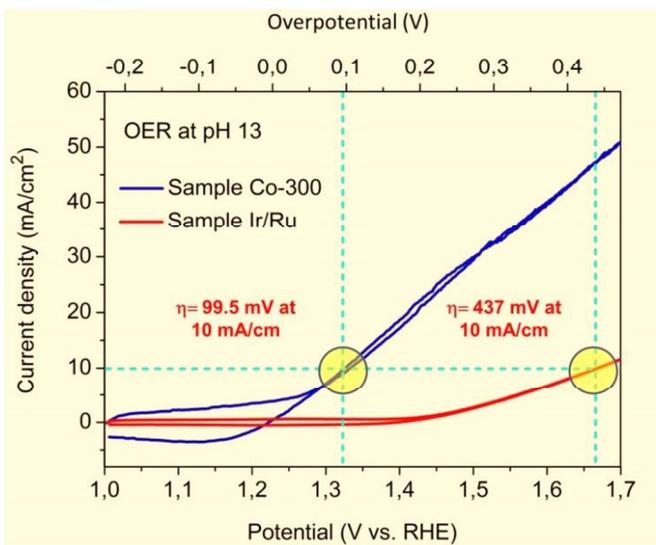
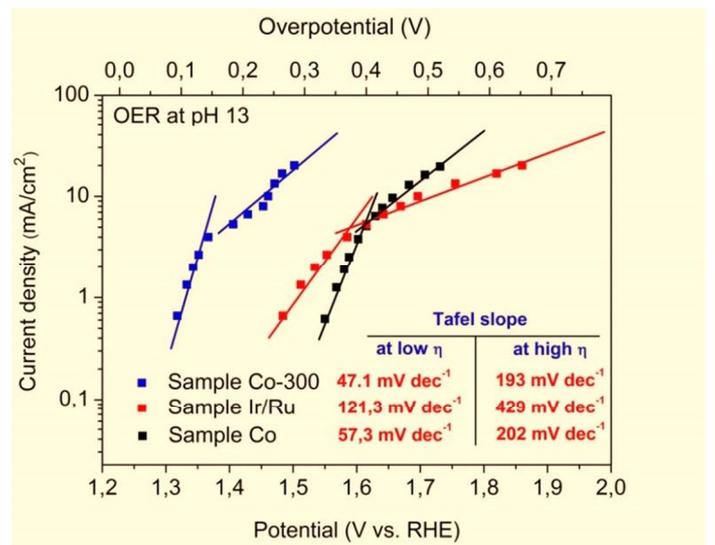
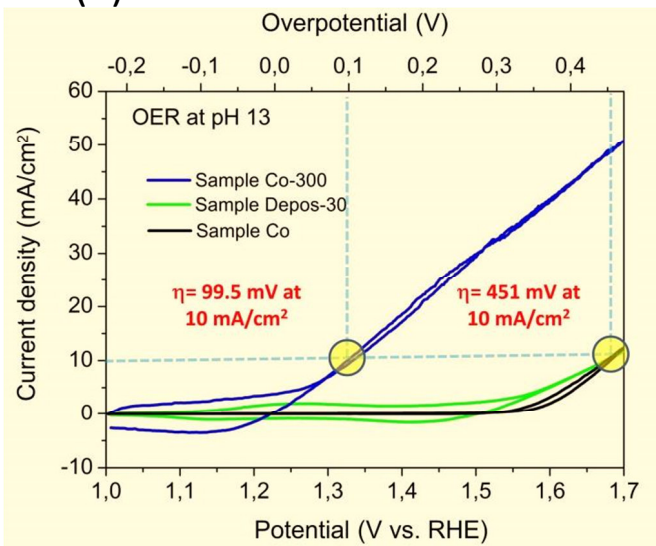

Figure 8. Comparison of the electrochemical OER properties of sample Co with sample Co-300- and Ir/Ru in 0.1 M KOH. Electrode area of all samples: 2 cm$^2$. **(a)** Cyclic voltammetric plots of sample Co/sample Co-300 based on 20 mV/s scan rate and 2 mV step size. **(b)** Long term chronopotentiometry measurement of sample Co-300 and chronopotentiometry measurement of sample Co performed at 10 mA/cm$^2$ current density. Average overpotential for the OER through 40000 s plot: 230 mV (Co-300). Average overpotential for the OER through 2000 s plot: 421 mV (Co). **(c)** Cyclic voltammetric plots of sample Co-300 and sample Ir/Ru based on 20 mV/s scan rate and 2 mV step size. **(d)** Tafel plots of samples Co, Co-300 and Ir/Ru based on 200 second chronopotentiometry scans at current densities 0.65, 1.33, 2, 2.66, 4, 5.33, 6.66, 8, 10, 13.3, 16.6 and 20 mA/cm$^2$. **(e)** Cyclic voltammetric plots of sample Co-300, Depos-30 and sample Co based on 20 mV/s scan rate and 2 mV step size.

As expected, sample Co-300 showed a much stronger current/voltage behavior compared to sample Ir/Ru at pH 13 (Figures 8c and 8d). The Tafel line of Co-300 was found to be much stiffer than the one of Ir/Ru or Co with a slope of 47.1 mV dec$^{-1}$ in the lower overpotential region (Ir/Ru:121.3 mV dec$^{-1}$, Co:57.3 mV dec$^{-1}$). Dual Tafel behavior was obtained throughout the sample series Co-300, Co and Ir/Ru with slopes up to 429 mV dec$^{-1}$ (Ir/Ru) in the higher potential region (Figure 8d). It turned out that via electrodeposition modified steel X20CoCrWMo10-9 (Depos-30) is also at pH 13 substantially weaker in supporting OER than Co-300. Depos-30 was proven to be just slightly better than untreated steel X20CoCrWMo10-9 (sample Co) as can be seen in Figure 8e, green curve.

**Conclusions**

The enlargement of the list of potential candidates suitable to efficiently and stable catalyze the anode half-cell reaction of the water splitting process is one of the most hotly contested subareas in "state of the art" energy science. Co$_3$O$_4$ has proven to be the most promising compound particularly when cleavage of water under neutral conditions is envisaged. Our efforts regarding the generation of steel-based-, outstandingly efficient, and stable OER electrocatalysts culminated in a composite consisting of X20CoCrWMo10-9 steel and a Co-oxide based ceramic. We investigated the performance of this composite material for anodic water splitting at pH 7 and pH 13. The efficiency at pH 7 was found to be higher than the one of any other material which has been tested by us ever and to the best of our knowledge it is substantially better than the capability of current noble- and non-noble OER electrocatalysts at pH 7 reported in literature so far. In addition, the material exhibited better OER key values at pH 13 than highly active-, surface modified AISI 304 steel on which we reported recently [26]. The outcome of our XPS- and FTIR studies on this new material can be reasonably explained by the formation of a $Co_3O_4$ containing "outer zone" on top of the steel. Upon a FIB-SEM study we could however show that the existence of $Co_3O_4$ solely does not explain the outstanding OER properties of sample Co-300. The existence of a classical substrate-layer structure with a substantial demarcation between periphery and substrate, typically characteristic for specimen achieved from electrodeposition based approaches (sample Depos-30) could not be proven in case of sample Co-300. A metal-ceramic transition can hardly be defined and an abrupt change in the composition is completely suppressed by this intrinsic "from within itself" formation of the X20CoCrWMo10-9 steel//$Co_3O_4$ composite. This particular microstructure is likely responsible for the unusual OER efficiency at pH 7 and at pH 13. Considering the fact that application options of $Co_3O_4$ are by far not restricted to electrocatalysis, we believe that our new X20CoCrWMo10-9 steel//$Co_3O_4$ composite holds big promises in terms of most diverse applications.

**Experimental Section**

*Preparation of the Co-300 samples-Electro oxidation of stainless steel at constant potential*

Samples with a total geometry of 70x10x1,5 mm were constructed from 1,5 mm thick X20CoCrWMo10-9 steel purchased from WST Werkzeug Stahl Center GmbH & Co. KG, D-90587 Veitsbronn-Siegelsdorf, Germany. Pre-treatment: Prior to each surface modification the surface of the metal was cleaned intensively with ethanol and polished with grit 400 SiC sanding paper. Afterwards the surface was rinsed intensively with deionized water and dried under air for 50 min at room temperature. The weight was determined using a precise balance (Sartorius 1712, 0.01 mg accuracy) prior to electro-activation. For the electro-oxidation a two-electrode set-up was used consisting of the steel sample as WE, and a platinum wire electrode (4x5 cm) used as CE. The WE (anode) was immersed exactly 2.1 cm deep (around 4.8 cm$^2$ geometric area), and the CE (cathode) was completely immersed into the electrolyte.

The electrolyte was prepared as follows: In a 330 mL glass beaker, 57.6 g (1.44 mol) of NaOH (VWR, Darmstadt, Germany) were dissolved under stirring and under cooling in 195 g deionized water. The solution was allowed to cool down to 23°C before usage. The anodization was performed in 200 mL of this electrolyte filled in a 300 mL glass beaker under stirring (450 r/min.) using a magnetic stirrer and a stirring bar (21 mm in length, 6 mm in diameter). The distance between WE and CE was adjusted to 6 mm. A power source (Electra Automatic, Vierssen, Germany) EA-PSI 8360-15T which allows to deliver a constant high current for longer operating time was used for the electrochemical oxidation. The procedure was carried out in current controlled mode. The current was set to 9 A according to ~1875 mA/cm$^2$ current density. The voltage varied during the electro-activation. At the beginning of the experiment it amounted to around 7.5 V but was reduced over a period of 300 min to around 4.6 V. These data proved to be reliably reproducible for all 10 replicates. If, however, for some reasons the decrease of the voltage is more abrupt, *i.e.* the voltage drops down to 4,6 V earlier, the activation procedure should be stopped when this value is reached. The electroactivation procedure should be stopped after 315 min at the latest. The temperature of the electrolyte increased within the first 30 min and reached a value of 323 K. Approximately 2.5 mL of fresh 6 M NaOH were added hourly to the electrolysis vessel in order to compensate the loss occurred due to evaporation. After 300 min of electro-activation the CE and the WE were taken out of electrolyte and rinsed intensively with tap water for 15 min and then with deionized water for a further 10 min. The used NaOH electrolyte was transferred quantitatively into a plastic bottle and was stored for further analysis. The counter electrode which was covered by a black layer, was immersed into 60 ml of 2 M HCl for 12 hours. The acidic solution was stored for further analysis. Prior to the electrochemical characterization the samples were dried under air at ambient temperature and the weight was determined upon a precise balance as described above. The sample preparation was repeated ten times, *i.e.* in total 11 samples of Co-300 have been prepared this way. The electrolyte of 5 activation procedures have been investigated via ICP OES, see supporting information.

*Preparation of Depos-30 samples*

The substrate with a total geometry of 70x10x1,5 mm was constructed from 1,5 mm thick X20CoCrWMo10-9 steel. Pre-treatment: Prior to each surface modification the surface of the metal was cleaned intensively with ethanol and polished with grit 800 SiC sanding paper. Afterwards the surface was rinsed intensively with deionized water and dried under air for 50 min at room temperature. The substrate was coated upon a twostep electrochemical approach consisting of (I) the electrodeposition of Co(OH)$_2$ and (II) the electrochemical oxidation of Co(OH)$_2$ to Co$_3$O$_4$. The electrodeposition of Co(OH)$_2$ was performed similar to a protocol described in literature by Jagadale *et al.* [57]

**Electrochemical Measurements**

A three-electrode set-up was used for all electrochemical measurements. The working electrode (WE) with a total geometry of 70x10x1,5 mm was constructed from 1,5 mm thick X20CoCrWMo10-9 steel, on which an apparent surface area of 2 cm$^2$ was defined by an insulating tape (Kapton tape). The IrO$_2$-RuO$_2$ sample (10 micrometer layer deposited on titanium) with a total geometry of 100x100x1.5 mm was purchased from Baoji Changli Special Metal Co, Baoji, China. An electrode area of 2 cm$^2$ was defined on the plate by Kapton tape. To avoid additional contact resistance the plate was electrically connected via a screw. A platinum wire electrode (4x5 cm geometric area) was employed as the CE, a reversible hydrogen reference electrode (RHE, HydroFlex, Gaskatel Gesellschaft für Gassysteme durch Katalyse und Elektrochemie mbH. D-34127 Kassel, Germany) was utilized as the reference standard, therefore all voltages are quoted against this reference electrode (RE). For all measurements the RE was placed between the working electrode and the CE. The measurements were performed in a 0.1 M KOH solution (VWR, Darmstadt, Germany) and in a pH 7 corrected 0.1 M KH$_2$PO$_4$/K$_2$HPO$_4$ solution (VWR, Darmstadt, Germany) prepared as follows: Aqueous solutions of 0.1 M K$_2$HPO$_4$ and KH$_2$PO$_4$ (VWR, Darmstadt, Germany) were mixed until the resulting solution reached a pH value of 7.0. The distance between the WE and the RE was adjusted to 1 mm and the distance between the RE and the CE was adjusted to 4-5 mm. All electrochemical data were recorded digitally using a Potentiostat Interface 1000 from Gamry Instruments (Warminster, PA 18974, USA), which was interfaced to a personal computer. Electrochemical measurements were carried out without any correction of the Ohmic voltage drop. With the exception of the measurements the Tafel plots are based on (Figure S2 new and…)

**Cyclic Voltammograms (CV)** were recorded in 90 mL of electrolyte (pH 7 corrected 0.1 M KH$_2$PO$_4$/K$_2$HPO$_4$, 0.1 M KOH respectively) in a 100 mL glass beaker under stirring (450 r/min.) using a magnetic stirrer (21 mm stirring bar). The scan rate was set to 20 mV/s and the step size was 2 mV for all OER related CV measurements. The potential was cyclically varied between 1 and 1.9 V vs. RHE for OER measurements at pH 7, between 1 and 1.7 V vs. RHE for OER measurements at pH 13, respectively.

**Chronopotentiometry scans** were conducted at a constant current density of 10 mA/cm$^2$ in 90 mL of electrolyte for measuring periods < 2000 s, in 800 mL of electrolyte for measuring periods ≥ 40000 s in a 100 mL respectively 1000 mL glass beaker. The scans were recorded under stirring (450 r/min.) using a magnetic stirrer (25 mm stirring bar) for measuring periods < 2000 s, using a magnetic stirrer (40 mm stirring bar) for measuring periods ≥ 40000 s respectively.

**Tafel plots**

Average voltage values for the Tafel plots were derived from 200 second chronopotentiometry scans at current densities of 0.65, 1.33, 2, 2.66, 4, 5.33, 6.66, 8, 10, 13.3, 16.6 and 20 mA/cm$^2$ for measurements carried out at pH 7 and pH 13. The arrangement of RE, WE and CE taken for recording the chronopontentiometry plots was as mentioned above (See paragraph *Electrochemical measurements*).

**Determination of Faradaic efficiency** (Figure 2, Table S2) was carried out in close accordance with the procedure described in Schäfer *et al.*, *Energy Envirnon. Sci*., **2015**, 8, 2685. Faradaic efficiency of OER was calculated by determining the dependence of the oxygen concentration in the electrolyte during the time of chronopotentiometry at constant current density of 5 mA/cm$^2$ and 10 mA/cm$^2$ in pH 7 corrected 0.1 M K$_2$HPO$_4$/KH$_2$PO$_4$ solution under stirring. The distance between RE and WE was adjusted to 1 mm, the distance between CE and RE was adjusted to ~7 mm. The volume of electrolyte was 2300 ml. The working compartment was completely sealed with glass stoppers before starting the chronopotentiometry at 0.07 mg O$_2$/l (j=10 mA/cm$^2$) and 0.06 mg O$_2$/l (j=5 mA/cm$^2$).

The results can be taken from **Figure** 2 and Table S2. The red line in **Figure** 2b corresponds to 100% Faradaic efficiency with a line equation: Y =7.20965 10$^{-4}$ x + 0.07 with y=Dissolved oxygen (mg/L); x=time (s) for the measurement performed at 10 mA/cm$^2$. The red line in **Figure** 2a corresponds to 100% Faradaic efficiency with a line equation: Y =3.60483 $10^{-4}$ x + 0.06 with y=Dissolved oxygen (mg/L); x=time (s) for the measurement performed at 5 mA/cm$^2$. The Faradaic efficiency amounted to 75.58% (10 mA/cm$^2$), 83.2% (5 mA/cm$^2$) respectively.

**XPS Spectroscopy (Samples Co-300-, Co)**

The XPS instrument used for measurements was a Multilab 2000 (ThermoVG Scientific) using a Mg X-ray source for the incident X-rays. Experiments were conducted at room temperature under ultra high vacuum conditions. Samples were etched using an Ar beam to remove adventitious carbon from the surface. Co samples required 5-65 minutes of etching depending on the element detected. No argon etching was applied to Co-300 samples.

**XPS Spectroscopy (Sample Depos-30)**

XPS measurements were performed using a Phoibos HSA 150 hemispherical analyzer equipped with standard Al Kα source with 0.3 eV full width at half-maximum. The measurements were recorded with the sample at room temperature. The spectra were calibrated using the carbon 1*s* line of adsorbed carbon ($E_B$ = 285.0 eV).

**FTIR Spectroscopy**

IR spectra were recorded on a Bruker Vertex 70 equipped with an ATR system *Golden Gate.* For the preparation of the powder, sample Co-300 was grinded by using grit 800 sanding paper till the black outer zone was completely removed from the specimen. The sanding paper loaded with material was subsequently put into double distilled water and the powder removed from the sanding paper by ultrasonic. The water was evaporated.

**Electron Microscopy**

Plan view SEM images of samples Co-300 and Depos-30 were taken on a Zeiss Auriga scanning electron microscope.

Cross sectional analysis (vertical plane imaging) of samples was realized by dual beam FIB (focused ion beam) –SEM technique. SEM images of the cross sections were taken on a Zeiss Auriga scanning electron microscope equipped with a Cobra FIB-column and Ga ion source using Feature Milling software module for modeling. The accelerating voltage was adjusted to 10 kV and the SEM images were acquired with a secondary electron- or back scatter detector.

**AFM Experiments**

AFM experiments. AFM measurements were done in semi-contact mode on a NT-MDT model NTEGRA Probe NanoLaboratory. The V-shaped cantilevers had nominal lengths of 140 µm, force constants of 25-95 N/m, and a resonance frequency of 242 kHz (within range of 200-400 kHz). The tip radius was 10 nm. The AFM images were processed by the software Nova Px.

Daniel M. Chevrier was supported by the NSERC CGS-Alexander Graham Bell scholarship and P.Z. acknowledges the NSERC Discovery Grant for funding. A special thank you is given to Andrew George (Dalhousie University, Department of Physics) for his technical assistance during XPS experiments. Mercedes Schmidt was supported by the European Research Council (ERC-CoG-2014; project 646742 INCANA).

Author contribution statement

HS had the idea to perform the experiments the manuscript is based on. He planned and performed the electrochemical measurements and all sample preparations. HS wrote the manuscript as well as the sup. Information. DMC and PZ planned, performed and evaluated the XPS measurements in Canada. KK. and JW planned, performed and evaluated the XPS measurements in Germany. Mercedes Schmidt performed and evaluated the AFM experiments. HS, UK and MS performed and evaluated the FIB/SEM investigations. DD planned and evaluated the ICP OES investigations. All authors have read the manuscript.